\magnification=\magstep1
\openup 1\jot
 
\input epsf

\font\list=cmcsc10
\font\cat=cmr7


\newcount\refno
\refno=0
\def\nref#1\par{\advance\refno by1\item{[\the\refno]~}#1}

\def\book#1[[#2]]{{\it#1\/} (#2).}

\def\annph#1 #2 #3.{{\it Ann.\ Phys.\ (N.\thinspace Y.) \bf#1}, #2 (#3).}
\def\annphl#1 #2 #3.{{\it Ann.\ Phys.\ (Leipzig) \bf#1}, #2 (#3).}
\def\apj#1 #2 #3.{{\it Ap.\ J.\ \bf#1}, #2 (#3).}
\def\cmp#1 #2 #3.{{\it Commun.\ Math.\ Phys.\ \bf#1}, #2 (#3).}
\def\cqg#1 #2 #3.{{\it Class.\ Quan.\ Grav.\ \bf#1}, #2 (#3).}
\def\grg#1 #2 #3.{{\it Gen.\ Rel.\ Grav.\ \bf#1}, #2 (#3).}
\def\jmp#1 #2 #3.{{\it J.\ Math.\ Phys.\ \bf#1}, #2 (#3).}
\def\mpla#1 #2 #3.{{\it Mod.\ Phys.\ Lett.\ \rm A\bf#1}, #2 (#3).}
\def\ncim#1 #2 #3.{{\it Nuovo Cim.\ \bf#1\/} #2 (#3).}
\def\npb#1 #2 #3.{{\it Nucl.\ Phys.\ \rm B\bf#1}, #2 (#3).}
\def\plb#1 #2 #3.{{\it Phys.\ Lett.\ \rm B\bf#1}, #2 (#3).}
\def\pla#1 #2 #3.{{\it Phys.\ Lett.\ \rm A\bf#1}, #2 (#3).}
\def\prb#1 #2 #3.{{\it Phys.\ Rev.\ \rm B\bf#1}, #2 (#3).}
\def\prd#1 #2 #3.{{\it Phys.\ Rev.\ \rm D\bf#1}, #2 (#3).}
\def\prl#1 #2 #3.{{\it Phys.\ Rev.\ Lett.\ \bf#1}, #2 (#3).}
\def\rpp#1 #2 #3.{{\it Rep.\ Prog.\ Phys.\ \bf#1}, #2 (#3).}
\def\sovj#1 #2 #3.{{\it Sov.\ J.\ Nucl.\ Phys.\ \bf#1}, #2 (#3).}

\overfullrule=0pt       

\def\zed{Z\hskip -3mm Z }
\def\half{\textstyle{1\over2}}

\def\spose#1{\hbox to 0pt{#1\hss}}
\def\lta{\mathrel{\spose{\lower 3pt\hbox{$\mathchar"218$}}
     \raise 2.0pt\hbox{$\mathchar"13C$}}}

\def\N{Nielsen}
\def\O{Olesen}

\rightline{gr-qc/9506054}
\rightline {DTP/95/37}
\rightline {SUSX-TH-95/73}
\vskip 0.6truecm

\centerline{\bf SMOOTH METRICS FOR SNAPPING STRINGS}
\vskip 0.6truecm

\centerline{ Ruth Gregory\footnote{$^\spadesuit$}{\sl Email:
rg10012@amtp.cam.ac.uk}}

\vskip 1mm

\centerline{\it Department of Applied Mathematics and Theoretical Physics,} 
\centerline{\it University of Cambridge, Silver St, Cambridge, CB3 9EW, U.K.} 

\centerline{\&}

\centerline{\it Centre for Particle Theory,} 
\centerline{\it Durham University, South Road,
Durham, DH1 3LE, U.K.}

\vskip 2mm
\centerline{and}
\vskip 2mm

\centerline{ Mark Hindmarsh\footnote{$^\clubsuit$}{\sl Email:
m.b.hindmarsh@sussex.ac.uk }}

\vskip 1mm

\centerline{\it 
School of Mathematical and Physical Sciences,}
\centerline{\it University of Sussex, Brighton BN1 9QH,UK}

\vskip 0.8cm
\centerline{\list abstract}
\vskip 2mm

{ \leftskip 5truemm \rightskip 5truemm

\openup -1\jot

We construct two possible metrics for abelian Higgs vortices with ends on 
black holes. We show how the detail of the vortex fields smooths out the
nodal singularities which exist in the idealized metrics. 
A corollary is that apparently topologically stable strings might
be able to  split 
by black hole pair production.  We estimate the rate per unit length 
by reference to related Ernst and C-metric instantons, concluding 
that it is completely negligible for GUT-scale strings.  
The estimated rate 
for macroscopic superstrings is much higher, although still 
extremely small, unless there is an early phase of strong coupling.

\openup 1\jot

\vskip 1 truecm\it PACS numbers: 04.40.-b, 04.70.-s, 11.27+d }

\vfill\eject
\footline={\hss\tenrm\folio\hss}

One of the basic tenets on which the cosmic string scenario [1-3]
of galaxy formation rests is that the only way strings decay
is via gravitational or particle radiation. The (straight)
Nielsen-Olesen vortex [4] has been taken to be stable, protected
by the topology of the vacuum manifold of the underlying
Abelian Higgs model. Such topological arguments have been 
assumed to be true even when gravity is included. While 
\N-\O~vortices seem to be stable to small perturbations
in Einstein Abelian Higgs [5], their stability against 
non-perturbative processes has not been confirmed.

Several recent papers have examined cosmic strings splitting under
various conditions [6-8], but only [8] actually considered the
splitting of a \N-\O~vortex. If true, these results might
have far reaching consequences for any large scale application of
cosmic strings if the decay rate is appreciable.
All of these arguments use metrics with conical deficits to find an 
instanton for decay and assume the conical deficit can be smoothed
out and replaced by a cosmic string. If a real vortex is to split
it is crucial to show that a real vortex can be woven into these
metrics, smoothly rounding off the conical deficit.

Evidence for the validity of replacing a conical deficit with a vortex was
presented in [9], where it was shown that the metric of Aryal Ford and
Vilenkin (AFV) [10], a black hole pierced by a conical deficit, 
could be considered
as the thin string limit of a vortex piercing a black hole. The main difference
between the AFV metric and the C-metrics, [11], used in [6-8]
is that the latter represents a
non-static process. From the technical point of view, the main difference is
that in the one case (AFV) we have an almost cylindrically symmetric situation,
whereas in the latter case, and also in the case of black holes in static
equilibrium, there is an asymmetry of the system in that the string locally
terminates on the event horizon. The string core no longer corresponds to
``$r=0$'' in an axial coordinate system, but only ``$r=0, z>z_0$''. 
Thus, although the results of [9] were suggestive, they were by no means
conclusive.

In this paper we demonstrate that the assumptions of [6-8] are
justified, in the sense that we show how 
to replace the conical singularity of both the uncharged
C-metric, as well as a static metric of  a string with ends, with an abelian
Higgs vortex. We will show that provided the mass of the black holes involved
is sufficiently large, the \N-\O~solution can be used to approximate the
field configuration, and the gravitational effect of the string will be shown
to smooth out the conical singularity into a `snub-nosed cone' [5]. A
consequence of this is that \N-\O~vortices need not be stable to
non-perturbative topology changing processes, and that strings might indeed
`split'.

We first consider the \N-\O~vortex and discuss the metrics on which it is 
supposed to sit. We then show how it smooths out any conical deficit by
calculating the back reaction of the vortex on the geometry.
After these more detailed considerations, we discuss some of the implications.
One is that topologically stable strings can break as indicated in [8]
by the nucleation 
of a pair of black holes, as the euclideanised version of the smoothed 
C-metric should represent an instanton by which a straight string can 
tunnel to the lorentzian configuration of black 
holes on Nielsen-Olesen vortices.
The tunneling rate can be estimated to be proportional to 
$\exp(-\pi m^2/\mu)$, where $m$ is the mass of the black holes 
and $\mu$ the string tension. We point out that as far as
strings at the Grand Unification scale are concerned, the tunneling 
rate is greatest for instantons where the 
black hole radius $2m$ is much smaller 
than the thickness of the string $\sim \mu^{-1/2}$, 
and so the process of splitting 
is really described by a different metric, which 
should resemble the Ernst [12] metric in the vicinity of the black holes.  
We estimate the rate, and find it still to be negligibly small for 
Grand Unified scale strings, but of possible significance for cosmic 
superstrings [13], whose string tension is much higher.  There is even  
the possibility of creating a population of primordial black holes, 
if the Universe ever went through a period where there was a significant 
population of macroscopic fundamental strings.

When 
\N-\O~strings were first conceived, they were meant to be
a realisation of the Nambu action, which allows for both closed and open
strings. The open strings would have to satisfy
certain boundary conditions, namely that the ends travel at the
speed of light. The reason that \N-\O~vortices were assumed not
to have ends is associated with the topology of the vacuum 
manifold. For future reference, the abelian Higgs lagrangian is
$$
{\cal L}[\Phi ,A_{\mu}] = D_{\mu}\Phi ^{\dagger}D^{\mu}\Phi -
{1\over 4}{ F}_{\mu \nu}{ F}^{\mu \nu} - {{\lambda  }\over 4 }
(\Phi ^{\dagger} \Phi - \eta ^2)^2,
\eqno (1)
$$
where $\Phi$ is a complex scalar field, $D_{\mu} = \nabla _{\mu} - ieA_{\mu}$
is the usual gauge covariant derivative, and
${ F}_{\mu \nu}$ the field strength associated with $A_{\mu}$.
We are using Planck units in which $G=\hbar=c=1$ and a mostly minus signature.
The vacuum manifold is $|\Phi|=\eta$, and therefore is a circle in the
complex plane. The \N-\O~vortex takes the form
$$
\Phi = \eta X_0(r)e^{i\phi}, \;\;\; ; \;\;\;
A_{\mu} = A_0(r)\nabla_{\mu}\phi,
\eqno (2)
$$
in cylindrical polar coordinates. Now, note that 
$$
\oint A_\mu dx^\mu = 2\pi A_0(r) \to {2\pi\over e}
\;\;\;\hbox{\rm outside the core}
\eqno (3)
$$
where the line integral is taken to be at constant $r$. More generally,
(i.e., if we are not in flat space, or do not have a straight string) the
presence of a vortex is indicated by the existence of closed loops in
space which lie totally in vacuo for which
$$
\oint A_\mu dx^\mu = {2\pi N\over e}, \;\;\; 
\eqno (4)
$$
for some $N\in$   \zed.
Then, Stokes' theorem is used to deduce the existence of a flux
tube crossing {\it any} surface spanning the loop -- and hence an
infinite or closed string. 

How then can a string have ends? One way, of course, is to embed 
the abelian Higgs model in a Yang-Mills-Higgs system with monopole solutions 
[1]. However, it was pointed out in [8] and [9] that as far as the topology 
of the fields is concerned, there is no obstruction to 
terminating the string on a black hole. The abelian gauge potential 
has to be defined in at least two patches on 2-spheres surrounding 
the black hole (just as in the Wu-Yang construction for magnetic monopoles 
[14]), so Stokes' theorem has to be used with care.  The 
spacetime is not topologically trivial, (the particular measure here being 
here the second cohomology class), therefore we cannot conclude that 
flux crosses every surface
spanning the loop -- only those surfaces deformable to one 
known to contain a vortex. 
Thus, depending on the actual spatial
topology, it is quite possible for a string to leave a neighbourhood
and thus effectively terminate as far as a local observer is concerned.

Such a situation occurs in the other metrics considered by Aryal, Ford 
and Vilenkin, namely the C-metrics [11] and a modification of the static
metrics considered by Israel and Khan [15]. A C-metric is an
axially symmetric solution to the Einstein equations which represents two black
holes uniformly accelerating apart. The force for this acceleration is provided
either by a conical excess,  a strut, between the holes, or alternatively by a
conical deficit,  a string, extending from 
each hole to infinity (or of course a 
combination of the two). The Israel-Khan metric represents two black holes
held in equilibrium by a strut, but can be readily modified to have two
strings extending to infinity.  The key observation about these metrics is that
the horizons of the  two black holes can be identified, forming a wormhole in
space [11,16].  The presence of this wormhole then provides a hole through which
the string can exit, thus it is not necessary  to consider charged black holes
and topologically unstable  strings, these uncharged metrics can directly
`swallow' a \N-\O~vortex. The basic idea then is to paint a vortex directly
onto the  metric, using the core to smooth out the conical deficit of the exact
metric. 

We will consider each metric in turn before showing that the
\N-\O~vortex can smooth out the conical deficit.
We will draw extensively
on the formalism of [9] and refer the reader there for calculational details.
There, the question  of using the vortex to place hair on the
black hole was considered, and it was shown that there was no
obstruction to having a vortex sit on the event horizon. In fact,
if the thickness of the vortex is less than the black hole radius 
($E=\sqrt{\lambda}\eta m>1$),
the \N-\O~solution was shown to be an excellent approximation to
the string fields and the vortex behaves almost as if the 
event horizon were not there. The fact that the event horizon
appears to cut the string from an external observers point of view
is readily explained by the open string boundary conditions. Recall 
that a Nambu string can end provided it is travelling at the speed of 
light. On the event horizon, the escape velocity {\it is} the speed of
light, so a `stationary' string sitting there is satisfying its
appropriate boundary conditions. 
For $E<1$, numerical results showed that the string was still relatively
unaffected by the black hole, although a slight pinching of the string
does occur.

We now show that for small $\mu$,  and string width much less than 
the black hole radius ($E\gg 1$), the \N-\O~solution solves the
abelian Higgs equations. We will neglect terms of order $\mu$ since these
correspond to the back reaction of the geometry on the abelian Higgs equations,
and can be accounted for via an iterative procedure.

First consider the 
C-metric which takes the form
$$
ds^2 = A^{-2}(x+y)^{-2}[{F(y) dt^2 - F^{-1}(y) dy^2 - G(x) 
d\phi^2/\kappa^2 - G^{-1}(x) dx^2]},
\eqno (5)
$$
where
$$
\eqalign{
G(x) &= 1-x^2-2mAx^3, \cr
F(y) &= -1 + y^2 - 2mAy^3. \cr
}
\eqno (6)
$$
Here, $m$ represents the mass of the black holes, and $A$ their acceleration. 
The factor $\kappa$ ensures that the axis between the black holes is regular,
and $\phi$ has periodicity $2\pi$. In the flat space limit, $A^{-1}$ represents
half the distance of closest approach, so if this metric were to represent a
string splitting, we would expect
$$
m\simeq \mu/A,
\eqno (7)
$$
where $\mu$ is the mass per unit length of the string. Let us write 
$x_1<x_2<x_3$ for the roots of $G$. Then, in order to obtain the correct 
signature, we must have $x_2<x<x_3$ and $-x_2<y<-x_1$.
The coordinates cover only one patch of the full spacetime corresponding to the
exterior spacetime of one accelerating hole up to its acceleration horizon, 
which is located at $y=-x_2$. The coordinate singularity at $y =
-x_1$ corresponds to the event horizon of the black hole. The conical 
deficit sits along $x=x_2$, while $x=x_3$ points towards the other black hole,  
which means that $\kappa= |G'(x_3)|/2$. The magnitude
of the deficit is given by 
$$ {\delta\over 2\pi} = 1 - \left | {G'(x_2) \over
G'(x_3)}\right | =  {x_3-x_2\over x_3-x_1}.
\eqno (8)
$$
Assuming $mA\ll 1$ in (6), the three roots $\{x_i\}$ are given by $\{ -{1/
2mA}, -1, 1\}$ so (8) requires $\delta=8\pi mA$. But $\delta=8\pi \mu$ for a
string, hence $\mu=mA$ in agreement with (7).

A more transparent form of the C-metric is obtained if we set
$$
{\bar t} = A^{-1}t \;\; , \;\; r = 1/Ay \;\; , \;\; {\rm and} \;\;\;
\theta = \int_x^{x_3} dx/\sqrt{G}
\eqno (9)
$$
when
$$
ds^2 = [1 + Arx(\theta)]^{-2}
\left [ (1-{2m\over r} - A^2r^2) d{\bar t}^2 - {dr^2 \over 
(1-{2m\over r} - A^2r^2)} - r^2 d\theta^2 - r^2 G d\phi^2/\kappa^2 
\right ].
\eqno (10)
$$
This is almost conformally equivalent to the Kottler[18], or Schwarzschild
de-Sitter metric as might be expected from the acceleration 
horizon and clearly shows that we reduce to the Schwarzschild metric 
in the limit $A\to 0$.

In order to solve the abelian Higgs equations, we rewrite the gauge field as
$$
A_\mu = {1\over e} (\partial_\mu \phi- P_\mu).
\eqno (11)
$$
Using the expression (2) for $\Phi$, the equations of motion are then
$$
\eqalignno{
\nabla _{\mu}\nabla ^{\mu} X
-  P_{\mu}P^{\mu}X + {\lambda   \eta  ^2\over 2} X(X^2 -1) &= 0,
&(12a) \cr
\nabla _{\mu}F^{\mu \nu} + 2 e^2 \eta^2 X^2 P^{\nu} &=0. \; 
&(12b) \cr }
$$
where $F_{\mu\nu} = \partial_\mu P_\nu - \partial_\nu P_\mu$ now. Noting that
in a normal spherically symmetric metric $X$ is a function of $r\sin\theta =
\sqrt{g_{\phi\phi}}$, we try $X = X_0(R)$, $P_\phi = P_0(R)$, where
$$
R = {\sqrt{\lambda G}\eta r  \over \kappa [1 + Arx(\theta)]},
\eqno (13)
$$
in the neighbourhood of $x=x_2$, the cosmic string.

Note that 
$$
r = {1\over Ay} \geq -{1\over Ax_1} = 2m,
\eqno (14)
$$
and
$$
{r\over 1+Axr} \simeq
{1\over A(x_2+y)} > {1\over A(x_2-x_1)} = {2m \over
1-2\mu},
\eqno (15)
$$
hence
$$
{\sqrt{\lambda }\eta r  \over [1+Arx(\theta)]} > 2E \gg 1.
\eqno (16)
$$
We therefore are interested in examining the equations of
motion over a range
$$
{\sqrt{G}\over \kappa} \leq 1/2E \ll 1.
\eqno (17)
$$
Thus
$$
\eqalign{
{\partial X \over \partial r} &= {\sqrt{\lambda G}\eta \over \kappa [1+Axr]^2}
X'(R) = \sqrt{\lambda }\eta X'(R) \times O(E^{-1}) \cr
{\rm and} \;\;\;\; \sqrt{G} {\partial X \over \partial \theta} 
&={\sqrt{\lambda G}\eta r\over \kappa [1+Axr]}
X'(R) \left [ {G'(x) \over 2} - {ArG\over (1+Axr)} \right ] 
=RX'(R) [ 1 + O(\mu E^{-2})]
\cr }
\eqno (18)
$$
Putting  this  form of $X$ into the X-equation of motion
yields
$$
X''[1+O(\mu)+O(E^{-2})]+{X'\over R}[1+O(\mu)+O(E^{-2})] + {XP^2\over R^2}
+ \half X(X^2-1) =0,
\eqno (19)
$$
which is indeed the \N-\O~equation for $X$ to the required order. The equation
for $P$ works similarly. Hence the \N-\O~vortex can be ``painted'' on to the
C-metric, in spite of its non-static nature.

Now consider the 
Israel-Khan metric which is given by
$$
ds^2 = e^{2\psi_o}dt^2 - e^{2(\gamma_o-\psi_o)}(dr^2+dz^2) 
- r^2 e^{-2\psi_o} d\phi^2,
\eqno (20)
$$
where, writing
$$
\eqalign{
\zeta &= \sqrt{\lambda}\eta z \;\;\;\;\;\; ;\;\;\; 
\rho = \sqrt{\lambda}\eta r
\cr
\zeta_1 &= \zeta - (L+{\half}E) \;\;\; ; \;\;\; 
\zeta_1' = \zeta - (L-{\half}E) \cr
\zeta_2 &= \zeta + (L-{\half}E) \;\;\; ; \;\;\; 
\zeta_2' = \zeta + (L+{\half}E) \cr
R_1^2 &= (\rho^2 + \zeta_1^2) \;\;\; {\rm etc.} \cr
E(i,j) &= R_iR_j + (\zeta_i\zeta_j + \rho^2) \;\;\; ; \;\;\; 
E(i',j) = R_iR_j + (\zeta'_i\zeta_j + \rho^2) \cr
 }
\eqno (21)
$$
we have
$$
\psi_o = {\half} \log \left [ {R_1 + R_1' - E\over R_1 + R_1' +E}\right ]
\left [{R_2 + R_2' - E\over R_2 + R_2' +E}\right ],
\eqno (22)
$$
and
$$
\gamma_o = {1\over 4} \log \left [ { E(1',1)^2E(1',2)^2E(1,2')^2E(2,2')^2 \over
E(1,1)E(1',1')E(1,2)^2E(1',2)^2 E(2,2)E(2',2')} \right ] - \log {4L^2-E^2\over
4L^2},
\eqno (23)
$$
where $E= \sqrt{\lambda}\eta m \gg 1$ represents (half) the black hole radius in
multiples of string width and $L$ represents the
separation of the black holes, also in units of string width. We can directly
find $L$ given the energy per unit length of the string, since the conical
deficit for $\zeta>L+\half E$ is
$$
\eqalign{
4\mu &= {\delta\over 2\pi} = 1-e^{-\gamma_o(\rho=0)} = {E^2 \over 4L^2} \cr
&\Rightarrow L = {E\over 4\sqrt{\mu}} \gg 1 \cr
}
\eqno (24)
$$

Now consider the string extending from
the upper black hole to infinity. We are then interested in a coordinate range
$\rho < \rho  e^{-\psi_o} \leq O(1)$ and
$\zeta>L+\half E$. Thus
$$
R_2+R_2' = 2(\zeta + L ) + O(\mu) + O(E^{-2})
\eqno (25)
$$
hence
$$
{R_2 + R_2' - E\over R_2 + R_2' +E} = 1 - {E \over \zeta  + L} +
O(\mu)
\eqno (26)
$$
Denote this quantity by $\log\psi_2$. Then $\psi_2 = O(\sqrt{\mu})$ and
$\psi_{2,z} = \sqrt{\lambda}\eta \times O(\mu/E)$. Similarly, $\gamma =
\gamma_{\rm sch} + \gamma_2$ where $\gamma_2$ is O($\sqrt{\mu})$ and has a
similarly suppressed variation. Thus the effect of the second black hole is to
multiply the Schwarzschild metric in the vicinity of the string core by an
extremely slowly varying factor. Therefore within the stated limits of the
approximation ($E\gg1$) the metric is Schwarzschild up to non relevant factors,
and the results of [9] can be used to conclude that the vortex equations can be
solved to the required order by the \N-\O~solution.

Now we turn to the gravitational back reaction. For this we use the 
canonical form of a general axisymmetric metric [19]
$$
ds ^2 = e ^{2 \psi }dt ^2 
-e ^{2( \gamma - \psi )}(dz ^2 + dr ^2 ) -
\tilde{\alpha}^2 e ^{-2 \psi } d \phi ^2.
\eqno (27)
$$
Note that although the Israel-Khan metric (20) is already in this form, the
C-metric is not. In order to make it so, one must perform the coordinate
transformation
$$
r={\sqrt{FG} \over \kappa A^2(x+y)^2} \;\;\; ; \;\;\;
z = -{ (1+xy+mA(x^3+2x^2y-y^3-2xy^2)) \over \kappa A^2(x+y)^2}
\eqno (28)
$$
In which case the metric comes into the canonical form of equation (20)
with
$$
\eqalign{
e^{2\psi_o} &= {F\over A^2(x+y)^2} \cr
e^{-2\gamma_o} &= {r^2A^4(x+y)^4 \over F} \left [ F \left (
{F'(y) \over 2F} - {2\over (x+y)} \right ) ^2 + G \left (
{G'(x) \over 2G} - {2\over (x+y)} \right ) ^2 \right ]
\cr }
\eqno (29)
$$
Although this appears rather messy, the key facts are that it does have the
canonical form, and that
$$
e^{-2\gamma_o(r=0)} = {G'(x_2)^2 \over 4\kappa^2 } = 
{G'(x_2)^2 \over G'(x_1)^2} .
\eqno (30)
$$

Now, writing
$$
\eqalign{
\rho &= \sqrt{\lambda}\eta r, \cr
\zeta &= \sqrt{\lambda}\eta z, \cr
\alpha &= \sqrt{\lambda}\eta\tilde{\alpha}. \cr
}
\eqno (31)
$$
the relevant Einstein equations from [9] are
$$
\eqalignno{
\alpha_{,\zeta\zeta} + \alpha_{,\rho\rho} &= -\epsilon \sqrt{-g}
({\hat T} ^\zeta_\zeta + {\hat T}^\rho_\rho) & (32a) \cr
(\alpha\psi_{,\zeta})_{,\zeta} + (\alpha \psi_{,\rho})_{,\rho} &=
{\half}\epsilon \sqrt{-g}({\hat T} ^0_0 - {\hat T} ^\zeta_\zeta
-{\hat T}^\rho_\rho - {\hat T}^\phi_\phi) & (32b) \cr
\gamma_{,\rho\rho} + \gamma_{,\zeta\zeta} &= - \psi_{,\rho}^2
- \psi_{,\zeta}^2 - \epsilon e^{2(\gamma-\psi)} {\hat T}^\phi_\phi
& (32c) \cr
}
$$
where $\epsilon=8\pi \eta^2$ represents the gravitational strength of the
string, and ${\hat T}^a_b = T^a_b/\lambda\eta^4$ is a normalised 
energy momentum tensor which is of order unity. The
combinations appearing above were all shown to be functions of 
$R=\rho e^{-\psi_o}$
in [9]. The slight alteration of the metric will not change this. The main
difference between the current calculation and the one presented in [9] is of
course the asymmetry. In [9] the Einstein equations were integrated out from
$r=0$ to obtain an asymptotically conical metric. Here we do not expect a
conical deficit over the full range of $\zeta$, 
therefore it is more appropriate to
place boundary conditions at the edge of the string rather than at its core.
That is, we take the metric perturbation to be non-zero only in the core.

Writing the metric functions as
$$
\gamma = \gamma_0 + \epsilon\gamma_1 \;\;\; \psi = \psi_0+\epsilon\psi_1 \;\;\;
\alpha = \rho( 1 + \epsilon \alpha_1)
\eqno (33)
$$
we expect $\alpha_1,\gamma_1,\psi_1 \to 0$ outside the core. Solving the
Einstein equations then gives
$$
\eqalign{
\alpha_1(R) &= - \int_R^\infty {1\over R^2} \int_0^R R^2[ {\hat T}^0_0 +
{\hat T}^\rho_\rho ]dR \cr
\psi_1(R) &= -{\half} \int_R^\infty R{\hat T}_\rho^\rho = 
{\half} \gamma_1(R)\cr
}
\eqno (34)
$$
over the relevant ranges of $\zeta$. Note that the perturbations fall to
zero outside the core\footnote{$^*$}{Note however that $\alpha_1$ actually
falls to zero as $O(E^{-1})/R$ [9]. Although this is outside the scope
of our present approximation, it is tempting to speculate that some relic
of this might remain when the calculation is continued to higher orders, 
and perhaps perturb the gravitational radiation present in the 
C-metrics [11], rather similar to the
way in which a string in a FRW universe radiates
C-energy as it preserves its proper radius [20].}, 
leaving the background metric (Israel-Khan or C),
but what of the interior metric? The real issue is what happens 
as $\rho\to 0$. Here there will be
a nodal singularity if
$$
\delta = 2\pi( 1 - \alpha'(0) e^{-\gamma(0)} ) 
= 2\pi ( 1 - (1+\epsilon\alpha_1(0))(1-\epsilon \gamma_1(0))
e^{-\gamma_o(0)} ) \neq 0.
\eqno (35)
$$
But
$$
\alpha_1(0) - \gamma_1(0) = - \int_0^\infty R{\hat T}_0^0dR = 4\mu/\epsilon
\eqno (36)
$$
and in each case,
the background metric was chosen to have a nodal
singularity (see equations (24) and (8)) of $8\pi\mu$ along this axis, i.e.,
$$
1- e^{-\gamma_o(0)} = 4\mu,
\eqno (37)
$$
hence $\delta=0$ to order $\mu$.
Thus the effect of the vortex is to smooth out the conical singularity 
giving a regular metric with  a
snub-nosed cone which is schematically depicted in figure 1.

\midinsert \hskip 5 truecm \epsfysize=8truecm
\epsfbox{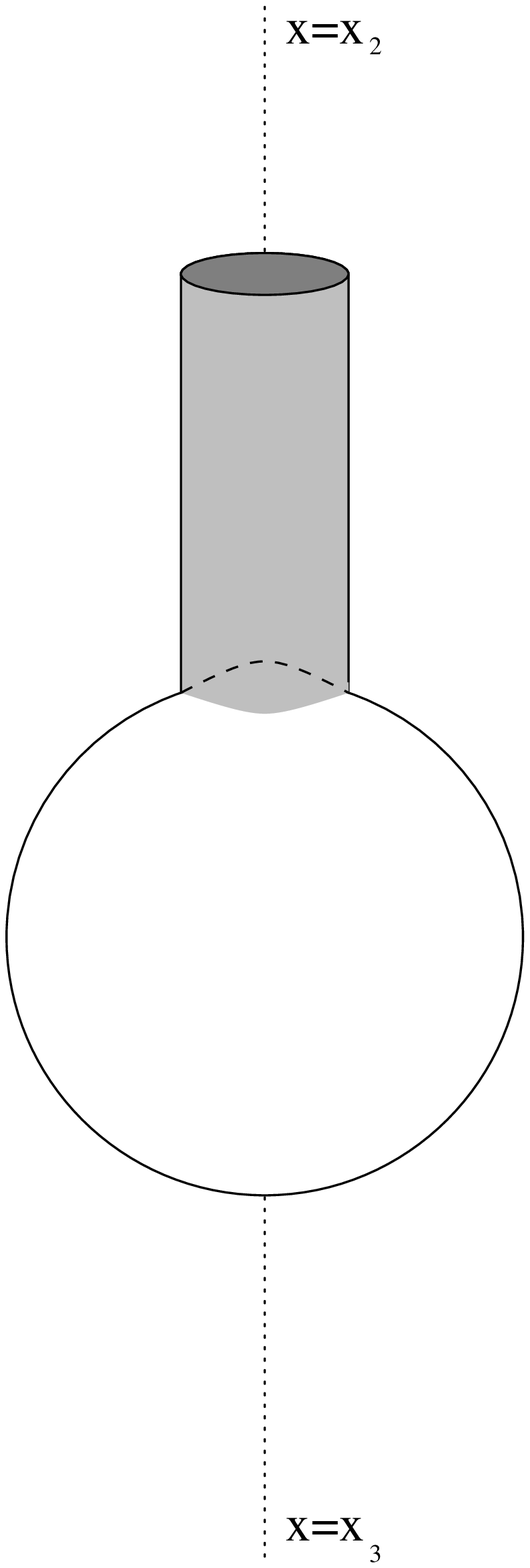} \medskip \hskip 2 truecm
\vbox{ \hsize=11.5 truecm
\noindent {\cat FIGURE (1): Pictorial representation of the geometry
of the vortex terminating on the black hole. }}
\endinsert

We have therefore shown that the \N-\O~solution can be used to construct 
regular metrics (in the sense of no nodal singularities) which represent
vortices which end on black holes either in  static equilibrium, or 
accelerating off to infinity. This latter metric
gives the appearance of a
cosmic string being eaten up by accelerating black holes. 
Whether or not such a
process can be used to destabilize cosmic strings depends on the action of the
corresponding instanton obtained by euclideanising the C-metric. Note that in a
model with vortices, we no longer require equality of the periodicities of
euclidean time at the event and acceleration horizon [17], since we can always
dress one or other horizon with an appropriate virtual string worldsheet
which can eat up any excess in periodicity of imaginary time [21], just as
the lorentzian string `eats up' the $\phi$-angle.
Indeed, since we are considering Einstein-abelian-Higgs a priori,
constructing such Euclidean vortices[21] to consume $\tau$-intervals in
the black holes geometry is a natural procedure to undertake. It
is arguably preferable to leaving a conical deficit there.

Euclideanizing (5) gives natural periodicities at the event and acceleration
horizons of
$$
\beta_e = {4\pi \over |G'(x_1)|} \;\;\; {\rm and} \;\;\;
\beta_a = {4\pi \over |G'(x_2)|}
\eqno (38)
$$
Combining this and equation (8) we can immediately see that 
demanding equality
of these periodicities requires $\mu=1/4$. Thus for small $\mu$,
$\beta_e\neq\beta_a$. In fact 
$$
\beta_e \simeq 8\pi\mu \gg \beta_a \simeq 2\pi
\eqno (39)
$$
So it appears that it is the acceleration horizon that must be
dressed if we wished to have a completely regular Euclidean section.
This would appear to imply that the action for such a process is
infinite, however, since calculations of the euclidean action are 
delicate [22], it would be premature to conclude that this must be the
case.

With this caveat in mind, let us assume that the action is finite, and
estimate it.
The instanton has the appearance of a two-dimensional plane
(the string worldsheet) with a disc, of radius $\rho$ say, removed from it. 
The difference between the action of such a configuration and that 
of the planar string worldsheet  is roughly
$$
I(\rho) = 2\pi\rho m - \pi\mu\rho^2.
\eqno (40)
$$
Extremizing this action with respect to $\rho$ gives the 
critical radius $\rho_{\rm c} =m/\mu$, and 
therefore the value of the action at the critical point is 
$$
I_{\rm c} = \pi m^2/\mu.
\eqno (41)
$$
However, in order to approximate the C-metric we had to assume 
that the vortex was much thinner than the width of the hole, or 
$$
m>{1\over\sqrt{\lambda}\eta} \sim \mu^{-1/2}.
\eqno (42)
$$
Thus, for Grand Unified strings, we are perforce considering instantons 
with enormous values of $I_{\rm c}$: in fact,
$$
I_{\rm c} \geq O(\mu^{-2}) \sim 10^{12}
\eqno (43)
$$
in agreement with the estimates of [8].

In order to reduce the action below this value we have to reduce the
size of the black hole below that of the string, and thus move 
away from the C-metric.  Recall that, apart from some pinching, the vortex
is essentially unaffected by the black hole horizon. This means that
in the limit $m \ll
{1/\sqrt{\lambda}\eta}$, the black holes can be contained entirely 
within the vortex, where the magnetic field is fairly uniform. 
In fact, near the centre of the vortex, the strength of the magnetic 
field is  $B = B_{\rm s} \simeq 2\pi/e m_{\rm  v}^2$, where 
$m_{\rm  v} = \sqrt{2} e \eta$ is the mass of the gauge field. 
We would expect this magnetic field to be able to nucleate 
magnetically charged black holes, and that the instanton 
describing this process to resemble locally the euclideanised Ernst 
metric [16,17] rather than the C-metric.  The Ernst metric [12], we recall, 
is an exact solution to the coupled Einstein-Maxwell field equations, 
describing a pair of oppositely charged black holes accelerating under 
the influence of a magnetic field.  However, the real metric 
describing a string splitting into a pair of small black holes 
in the Einstein-Abelian-Higgs system must confine the magnetic flux 
to within a distance $m_{\rm v}^{-1}$ of the axis of symmetry, and 
thus we would expect to recover the C-metric at spacelike separations 
much greater than this. 
At intermediate scales, the flux lines emanating from the black hole
will get swept up and around into the confined vortex extending to
infinity, as in figure 2.
 
\midinsert \hskip 5 truecm \epsfysize=8truecm
\epsfbox{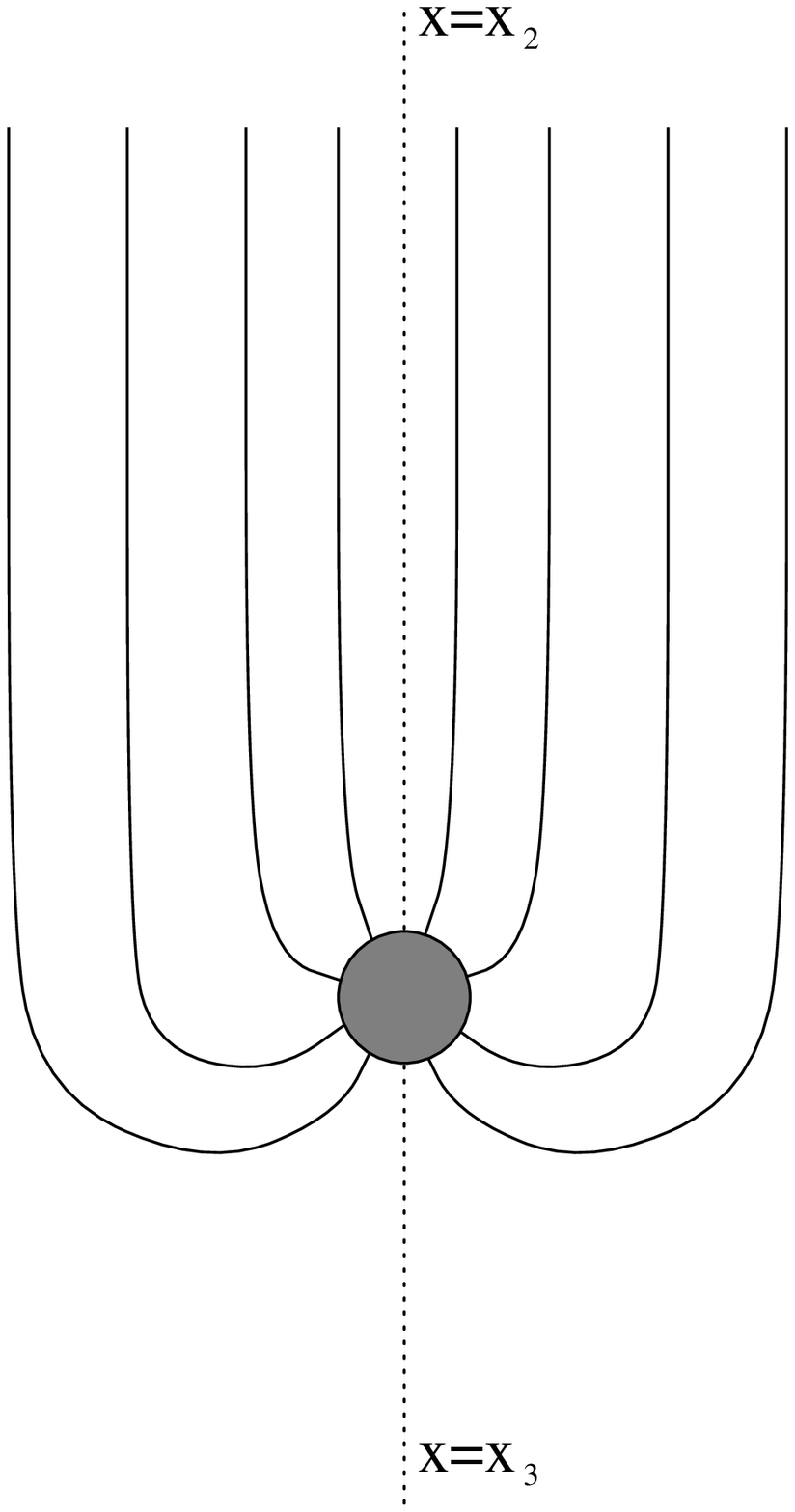} \medskip \hskip 2 truecm
\vbox{ \hsize=11.5 truecm
\noindent {\cat FIGURE (2): A representation of the nucleation of
a small black hole within a thick vortex.}}
\endinsert

The extra action for the creation, separation, and annihilation 
of a pair of virtual black holes in the background field $B$ 
can be estimated as
$$
I(\rho) = 2\pi m \rho - \pi q B \rho^2,
\eqno (44)
$$
where $q=2\pi/e$ is the magnetic charge of the black hole. This 
is extremised at $\rho_{\rm c} = m/qB$, so 
$$
I_{\rm c} = \pi m^2 / qB.
\eqno (45)
$$
In the background supplied by the core of the vortex, we find 
(using $\mu = 2\pi\eta^2$, which is only strictly true 
in when $\lambda = e^2/2$) 
$$
I_{\rm c} = m^2/2\mu.
\eqno (46)
$$
This is different by a factor of $2\pi$ from the C-metric 
action extrapolated beyond its domain of validity [6].
The minimum possible action is obtained when the hole is extremal, 
$m=q$, for which
$$
I_{\rm c} = 2\pi/e^2\mu,
\eqno (47)
$$
which for Grand Unified strings is of order $10^7$.

With this action we can estimate the rate per unit length of string 
$\gamma$ for the breaking process as
$$
\gamma \sim M^2 \exp(-2\pi/e^2\mu),
\eqno (48)
$$
where $M$ is a mass scale in the problem.  This can only be 
calculated by evaluating the determinants of small fluctuations 
in the instanton background, which is beyond the scope of the present
paper. However, we can estimate the prefactor by drawing 
on the known rate per unit volume for particle creation in 
a uniform electric field $E$ [23], which 
is $(e^2E^2/8\pi^3)\exp(-\pi m^2/eE)$.  Since the field is 
confined to a tube of area $m_{\rm v}^{-2}$,
the rate should be approximately
$$
\gamma \sim e^{-4}\mu\exp(-2\pi/e^2\mu).
\eqno (49)
$$
This rate is utterly negligible for GUT-scale strings.  Even if 
the Universe was crammed with strings, so that they begin to 
overlap, the rate of pair creation would still be only 
of order $\mu^2 t^3 \exp(-2\pi/e^2\mu) \sim 10^{200}\exp(-10^6)$.

The best we can hope to do is to put 
a macroscopic superstring at the conical deficit of the C-metric.
Of course, we should in that case solve the low-energy superstring 
field equations, including the dilaton and the 
antisymmetric tensor field. (This would be very similar
to the calculation of [17], however with an axion, rather than 
electromagnetic, field.)  However, it is known that the 
metric around a superstring is also conical [24], 
so it is not at all 
inconceivable that an exact solution (and an associated instanton) 
similar to the C-metric exists in this theory. 

The mass per unit length of a macroscopic superstring is 
$g^2/32\pi^2$ [25], where $g$ is the Grand Unified gauge coupling, which 
is of order $10^{-3}$ at the GUT scale.  
We could not reasonably expect the black 
holes to have less than the Planck mass, and so the exponential 
factor is of order $\exp(-10^3)$.  The entire history of the 
visible universe occupies only about $10^{240}$ Planck units of 
spacetime volume, and so the splitting process is negligible 
even for superstrings, unless there is a period where 
the gauge coupling constant is large.  In that case, there might be  
a relic population of primordial black holes left behind by an 
early phase of breaking superstrings.

To summarise: we have demonstrated that it is possible to replace
the conical singularities of the C-metrics and the Israel-Khan metric
with a vortex solution of the abelian Higgs model by calculating the
gravitational back reaction to linear order in $\mu$, the energy per
unit length of the string. We consider the implications for the splitting
of cosmic strings and argue that if the decay does proceed by 
an instanton in the euclidean theory, then it will in any case be 
suppressed by a ludicrously
large factor even for GUT strings. The tunnelling process for cosmic
superstrings, which have a yet larger string tension, is still extremely small. 
This
suggests that although it is of great interest that otherwise topologically
stable strings might be unstable, it is probably of no relevance to
practical applications.
 
{\it Note added in proof.} We thank G.Horowitz for pointing out a problem
with our suggestion of splitting superstrings, namely, that the
axion field has a conserved topological charge. We have left this argument
in case the reader can find a resolution. We would also like to thank
Simon Ross for pointing out an error in the initial version of the
manuscript.

\medskip
\noindent{\bf Acknowledgements}

We would like to thank Ana Achucarro and Konrad Kuijken for 
conversations.
This work is supported by PPARC Advanced Research Fellowships 
B/93/AF/1716 (R.G.) and B/93/AF/1642 (M.H.).

\bigskip

\noindent{\bf References}
\frenchspacing

\nref
M.B. Hindmarsh and T.W.B. Kibble, \rpp 58 477 1995.

\nref
A. Vilenkin and E.P.S. Shellard, {\it Cosmic strings and other 
Topological Defects} (Cambridge Univ. Press, Cambridge, 1994).

\nref
R.H. Brandenberger, {\it Modern Cosmology and Structure Formation}
astro-ph/9411049.

\nref
H.B. Nielsen and P. Olesen, \npb 61 45 1973.
 
\nref
R. Gregory, \prl 59 740 1987.
 
\nref
S.W. Hawking and S.F. Ross, {\it Pair production of black holes on cosmic
strings}, gr-qc/9506020.

\nref
R. Emparan, {\it Pair creation of black holes joined by cosmic 
strings}, gr-qc/9506025.

\nref
D. Eardley, G. Horowitz, D. Kastor and J. Traschen, {\it Breaking 
cosmic strings without monopoles}, gr-qc/9506041.
 
\nref
A. Achucarro, R. Gregory and K. Kuijken, {\it Abelian Higgs Hair for Black 
Holes},  gr-qc/9505039.

\nref
M. Aryal, L. Ford and A. Vilenkin, \prd 34 2263 1986.
 
\nref
W. Kinnersley and M. Walker, \prd 2 1359 1970.
 
\nref
F.J. Ernst, \jmp 17 515 1976.

\nref
E. Witten, \plb 153 243 1985.

\nref
T.T. Wu and C.N. Yang \prd 12 3845 1975, \npb 107 365 1976.

\nref
W. Israel and K.A. Khan, \ncim 33 331 1964.
 
\nref
D. Garfinkle and A. Strominger, \plb 256 146 1991.

\nref
H.F. Dowker, J.P. Gauntlett, D.A. Kastor and J. Traschen, \prd 49 2909 1994.

\nref
F.Kottler, \annphl 56 401 1918. 

\nref
J.L. Synge, \book Relativity: The General Theory [[North Holland, Amsterdam, 
1960]]

\nref
R.Gregory, \prd 39 2108 1989.

\nref
H.F. Dowker, R. Gregory and J. Traschen, \prd 45 2762 1992.

\nref
S.W.Hawking and G.T.Horowitz, {\it The Gravitational Hamiltonian, action,
entropy and surface terms},  gr-qc/9501014.

\nref
C. Itzykson and J-B. Zuber, \book Quantum Field Theory 
[[McGraw-Hill, New York, 1980]]

\nref
A. Dabholkar, G. Gibbons, J.A. Harvey and F. Ruiz Ruiz, \npb 340 33 1990.

\nref
N. Turok, \prl 60 549 1988.

\bye